\documentclass[10pt,emulateapj,apj]{emulateapj}
\shorttitle{Genus Topology of LRGs}
\shortauthors{Gott et al.}
\begin{document}
\title{3D Genus Topology of Luminous Red Galaxies}
\author{J. Richard Gott III\altaffilmark{1}, 
Yun-Young Choi\altaffilmark{2}, 
Changbom Park\altaffilmark{3}, 
and Juhan Kim\altaffilmark{4}}
\altaffiltext{1}{Department of Astrophysical Sciences, Peyton Hall, Princeton University, Princeton, NJ 08544-1001, USA}
\altaffiltext{2}{School of Space Research, Kyung Hee University, Yongin,
Gyeonggi 446-701, Korea; yychoi@kias.re.kr}
\altaffiltext{3}{Korea Institute for Advanced Study, Dongdaemun-gu, Seoul 130-722, Korea; cbp@kias.re.kr}
\altaffiltext{4}{Canadian Institute for Theoretical Astrophysics, 
University of Toronto, 50 St. George Street, Toronto, ON M5S 3H4, Canada}

\begin{abstract}
We measure the 3D genus topology of large scale structure using 
Luminous Red Galaxies (LRGs) in the Sloan Digital Sky Survey
and find it consistent with the
 Gaussian random phase initial conditions expected from the simplest scenarios
of inflation.  This studies 3D topology on the largest scales ever obtained.
 The topology is sponge-like.  We measure topology in two volume-limited 
samples: a dense shallow sample studied with smoothing length of 21$h^{-1}$Mpc, 
and a sparse deep sample studied with a smoothing length of 34$h^{-1}$Mpc.  
The amplitude of the genus curve is measured with 4\% uncertainty. Small 
distortions in the genus curve expected from non-linear biasing and gravitational
effects are well explained (to about 1-$\sigma$ accuracy) by 
$N$-body simulations using a subhalo-finding technique to locate LRGs. 
This suggests the formation of LRGs is a clean problem that can be modeled well
without any free fitting parameters.   This bodes well for using LRGs to
measure the characteristic scales such as the baryon oscillation 
scale in future deep redshift surveys.  
\end{abstract}

\keywords{large-scale structure of universe -- cosmology: observations -- 
methods: numerical}

\section{Introduction}

Topology of large-scale structure in the universe 
has been studied over the years through analyses of 
large-scale distribution of galaxies in three dimensions. 
The genus statistic was used to characterize the topology
(Gott, Melott, \& Dickinson 1986; Hamilton, Gott, \& Weinberg 
1986; Gott, Weinberg, \& Melott 1987; Gott, et. al. 1989; 
Vogeley, et. al. 1994; Park, Kim, \& Gott 2005$a$; Park et. al. 2005$b$). 
Measuring the genus as a function of density (Gott et al. 1986, 1987)
allows us to compare the topology observed 
with that expected for Gaussian random phase initial conditions, as predicted 
in a standard big bang inflationary model (Guth 1981; Linde 1983) where 
structure originates from random quantum fluctuations in the early universe. 


At large scales where fluctuations are still in the linear regime, 
growing in place without changing topology, 
topology at the median density level 
is expected to be sponge-like (Gott et al. 1986, 1987).  
Small deviations from the random phase curve give important information 
about biased galaxy formation and non-linear gravitational clustering as shown 
by perturbation theories and large $N$-body simulations 
(Matsubara 1994; Park, et al. 2005$a$).  
Previous models of galaxy clustering suggested either a meatball topology, 
isolated clusters growing in a low density connected background 
(Press \& Schechter 1974; Soneira \& Peebles 1978), or a Swiss cheese topology, 
isolated voids surrounded on all sides by walls (Einasto, 
Joeveer, \& Saar 1980).  
But studies of many observational samples 
have shown in every case a sponge-like median density contour as expected 
from inflation (Gott et al.  1986, 1989, 2008; 
Moore et al. 1992; Vogeley et al. 1994; Canaveses et al. 1998; Hikage
et al. 2002, 2003; Park et al. 2005$b$).

We have previously measured the 3D topology of the nearby Main galaxies 
(Strauss et al. 2001) of the Sloan Digital Sky Survey (York et al. 2000;
Stoughton et al. 2002; Gunn et al. 2006) 
data at smoothing lengths ranging from 3.5 to 11
$h^{-1}$ Mpc (see Park et al. 2005$b$ for details). 
The overall agreement with the best-fit 
Gaussian random phase genus curve and the mock surveys in a $\Lambda$CDM
is remarkable, strongly supporting the predictions of inflation.  
Small deviations from 
the Gaussian random phase distribution are expected because of non-linear 
gravitational evolution and biased galaxy formation, and these can now be 
observed with sufficient accuracy to do model testing of galaxy formation 
scenarios.
In this paper we will analyze the 3D distribution
of the luminous red galaxys (LRGs; Eisenstein et al. 2001) 
observed by the SDSS to study the genus topology 
at very large scales.  The results will be compared with those
from the nearby galaxies and with those from N-body simulations
of the inflation-based $\Lambda$CDM universe.

\section{Luminous Red Galaxies}
The LRG galaxies in the SDSS have already proven useful 
in refining cosmological parameters (Tegmark et al. 2006). 
Measuring the genus topology of the LRGs is of particular importance because it 
allows us to observe topology on the largest scales, and because the LRGs are 
expected to play an important role in characterizing dark energy,
i.e. through the ratio of dark energy pressure to energy density
(cf. Bessett et al. 2005;
Eisenstein et al. 2005; and Perceival et al. 2007 for a review).


We use the LRG sample from the SDSS.
The sky coverage of the sample is similar to that of SDSS data Release 5
(Adelman-McCarthy et al. 2006). The angular selection function is based
on the spherical polygon description of LSS-DR4plus catalog
(Blanton et al. 2003, 2005).
To maximize the volume-to-surface ratio
we trim the DR4plus sample (see Figure 1 and the text of Park et al. [2007]).
These cuts leave 68,203 LRGs over about $4464$ deg$^2$ in the survey region.  

We constructed two volume-limited subsamples of LRGs in
passively evolved luminosity and redshift ranges.
The subsamples are $-21.2<M_{g,0.3}<-23.2$ with $0.2<z<0.36$, and
$-21.8<M_{g,0.3}<-23.2$ with $0.2<z<0.44$, where we have applied $k$-corrections
and passively evolved the galaxies to a fiducial redshift of 0.3.
We label them ``SHALLOW''and ``DEEP'', respectively. 
Both subsamples are selected to reduce the artifacts caused by the LRG selection cuts. The resulting comoving number density of LRGs are approximately constant. (see Zehavi et al. 2005 for details).
The sample DEEP (SHALLOW) contains 
32,067 (14,968) galaxies. The absolute magnitude and 
redshift limits of these samples are illustrated in 
Figure 1.  We will measure the genus topology of the DEEP (SHALLOW) sample 
with a smoothing length of 21 (34) $h^{-1}$Mpc.
The smoothing lengths chosen are large enough to 
avoid unwanted shot noise effects (c.f. Gott, Melott \& Dickinson 1985) 

\section{$N$-body simulation}

We completed a $2048^3$ particle $4915.2 h^{-1}$Mpc size
Cold Dark Matter Simulation adopting the WMAP 3-year parameters
(Spergel et al. 2007); 
$\Omega_m=0.238, \Omega_{\Lambda}=0.762, \Omega_b=0.042, n_s=0.958, \sigma_8=0.761$, 
and $h=0.732$,  where $\Omega_m, \Omega_{\Lambda}, \Omega_b$ are 
the density parameters
due to matter, cosmological constant, and baryon, respectively.
$n_s$ is the slope of the power spectrum at large scales, and
$\sigma_8$ is the rms density fluctuation at $8 h^{-1}$Mpc scale.
The simulation is run by a PM+Tree N-body code (Dubinski et al. 2004)
and has a force resolution of 0.24 $h^{-1}$Mpc.

We identify the most massive dark matter subhalos as LRGs.
The subhalos are the gravitationally bound halos not 
subject to tidal disruption (Kim \& Park 2006), and having more than
30 member particles.  
To compare with the observational data we saved the particle positions
and velocities along the past light cone of 12 well-separated observers
who are making mock SDSS DR4plus LRG surveys.
The subhalos are found on the past light cone surface.
The minimum halo masses 
matching the observed number densities of the SHALLOW and DEEP LRG
samples are found as a function of redshift 
and applied to the subhalos data on the past line cone 
when mock surveys are made.
In each mock survey we pick the most massive subhalos 
when the mean number density of the LRGs is matched.
At small non-linear scales the genus depends sensitively on the method 
by which galaxies are identified within the simulation.
Park et al. (2005a) and Gott et al. (2008) have extensive studies on this 
question.

\begin{figure}
\epsscale{1.}
\plotone{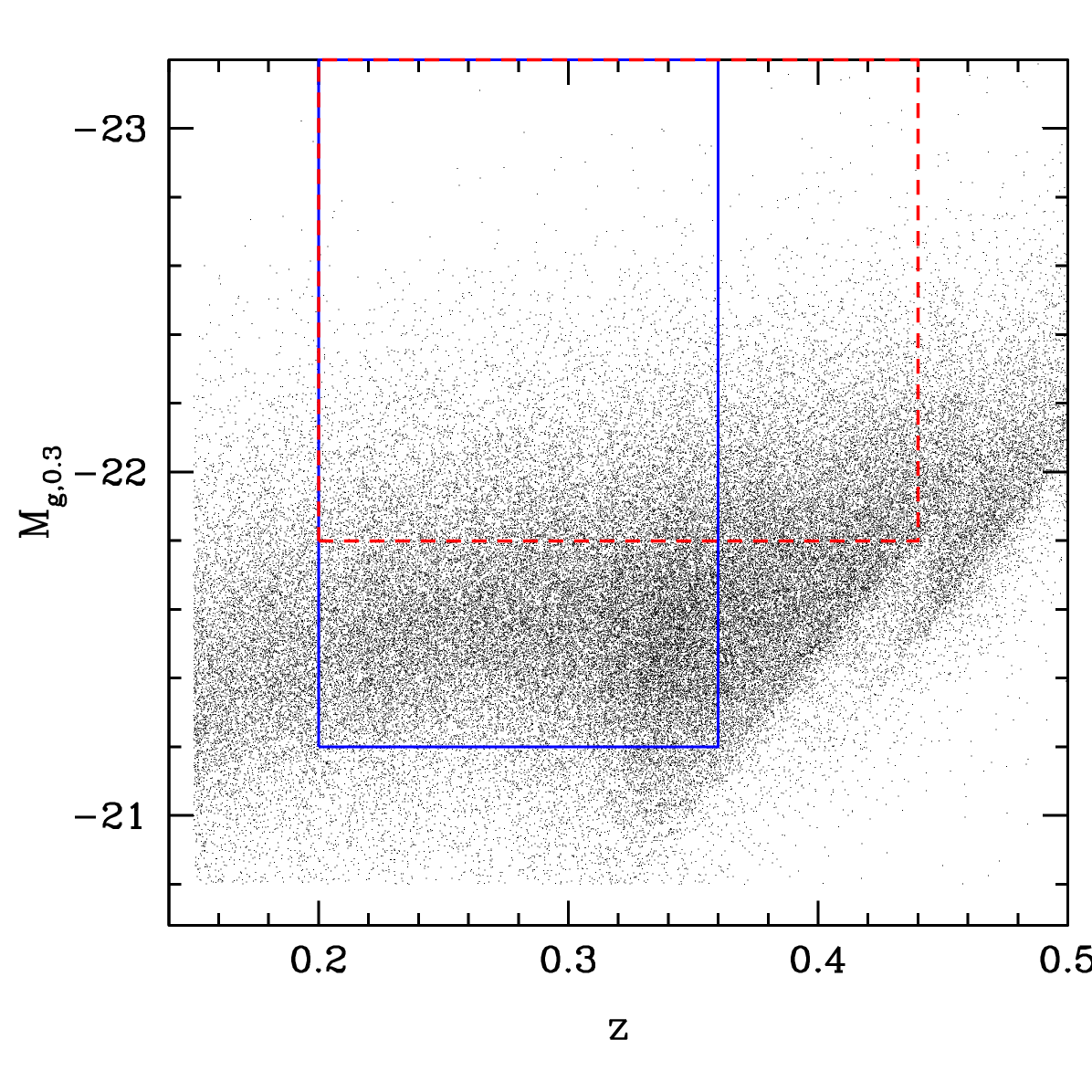}
\caption{
Distribution of Luminous Red Galaxies of the SDSS. Solid boundary lines define
the SHALLOW sample, and the dashed lines define the DEEP sample. }
\end{figure}


\section{The Genus and Related Statistics}

To measure the genus
we smooth the LRG number density distribution with a Gaussian smoothing 
ball of radius $R_G$. 
To remove the small radial variation of LRG number density 
in our samples, we divide the density by the radial selection function 
in the Appendix of Zehavi et al. (2005).
Then the
iso-density contour surfaces of the smoothed galaxy density distribution
are searched to calculate the genus. The contour surfaces are labeled by $\nu$, 
which is related with the volume fraction $f$ on 
the high density side of the density contour surface by
\begin{equation}
f = {1\over\sqrt{2\pi}}\int_\nu^\infty e^{-x^2/2} \,dx.
\end{equation}
The $f = 50\%$ contour 
corresponds to the median volume fraction contour ($\nu = 0$). 
For Gaussian random phase initial conditions the genus curve is:
\begin{equation}
\label{eq:GRgenus}
g(\nu) = A(1 - \nu^2)e^{-\nu^2/2},
\end{equation}
where the amplitude $A = (\langle k^2 \rangle /3)^{3/2}/2\pi^2$
and $\langle k^2 \rangle$ is the average value of $k^2$ in the
smooth power spectrum
(Hamilton, Gott, \& Weinberg 1986; Doroshkevich 1970)

The shape of the genus curve can be parameterized by several variables. 
First, there is the amplitude of the genus curve as measured by fitting the 
amplitude of the best fitting Gaussian random phase curve from Equation 2. 
This gives information about the power spectrum and phase correlation
of the density fluctuation.  Deviations of the genus 
curve from the theoretical random phase case can be quantified 
by the following three variables:
\begin{equation}
\Delta\nu = \frac{\int_{-1}^1 g(\nu)\nu\,d\nu}{\int_{-1}^1
  g_{\mbox{rf}}(\nu)\,d\nu}, ~~~
\end{equation}
\begin{equation}
A_V = \frac{\int_{-2.2}^{-1.2}g(\nu)\,d\nu}
{\int_{-2.2}^{-1.2}g_{\mbox{rf}}(\nu)\,d\nu}, ~~~
A_C = \frac{\int_{1.2}^{2.2}g(\nu)\,d\nu}
{\int_{1.2}^{2.2}g_{\mbox{rf}}(\nu)\,d\nu},
\end{equation}
where $g_{\rm rf}({\nu})$ is the genus of the best-fit random phase curve 
(Eq. 2). Thus $\Delta\nu$ measures any shift in the central portion of the 
genus curve.  The Gaussian curve (Eq. 2) has $\Delta\nu = 0$.  
A negative value of $\Delta\nu$ is called a ``meatball shift'' 
caused by a greater prominence of isolated connected high-density 
structures which push the genus curve to the left. 
$A_V$ ($A_C$) measure the observed number of voids (clusters) 
relative to those expected from the best-fitting Gaussian curve. 


\section{Results}

\begin{figure}
\epsscale{1.1}
\plotone{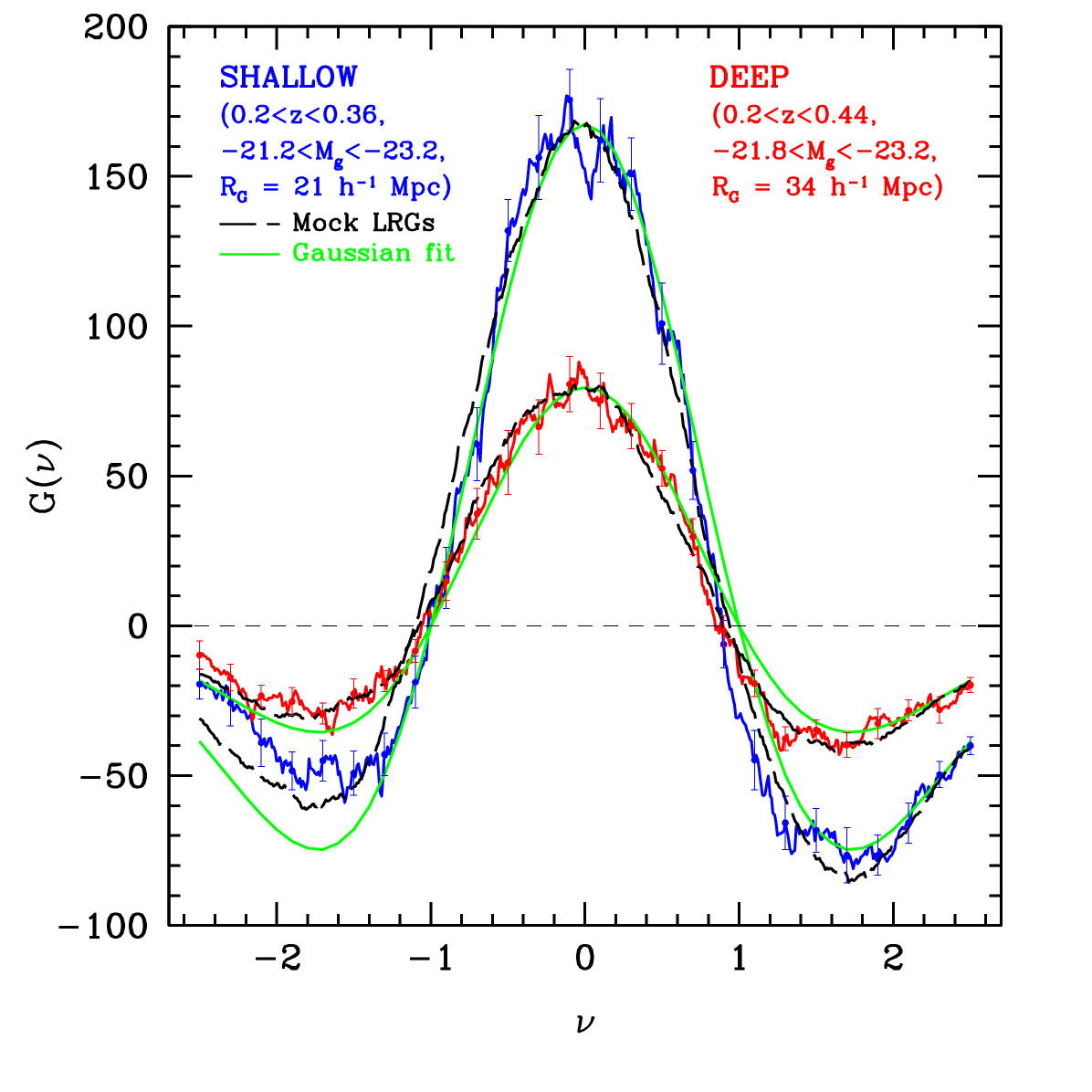}
\caption{
Genus curves for observed (two noisy curves) and simulated (two long-dashed curves) 
Luminous Red Galaxies. Gaussian fits (solid lines) are also shown. 
The genus curves with higher
amplitudes are for the SHALLOW sample smoothed with 
$R_G=21 h^{-1}$Mpc, and those with
lower amplitudes are for the DEEP sample with $R_G=34 h^{-1}$Mpc.
}
\end{figure}

We have measured the 3D genus topology of the LRGs in the SHALLOW and 
DEEP samples with the smoothing lengths described above.  The genus curves 
are shown in Figure 2.
The maximum value of the genus curve reaches about 170, and the amplitude
is measured with only 4\% error at $21 h^{-1}$Mpc scale. 
This compares with the previously smallest
error of 9.4\% at 5 $h^{-1}$Mpc scale for the genus obtained from the Main
Galaxy Sample of the SDSS (Park et al. 2005$b$).
The data for each sample is shown as the jagged line.  
Each genus curve is 
compared with the best-fit Gaussian random phase curve from Equation 2 
(smooth solid lines).  The observational data agree well with the Gaussian
curves, but there are small differences between them due to the non-linear
biasing and gravitational effects (Park et al. 2005$a$).



The dashed lines show the mean of 12 mock SDSS Surveys 
constructed from our $2048^3$ particle simulation. 
Although our two volume-limited samples contain galaxies typically with different absolute magnitude and redshift, the genus curves measured from them show a very good overall agreement. Moreover,
there is a remarkable agreement between observations and the mock surveys.
The number of voids (the depth of the valley at $\nu< -1$), 
is less than the number of clusters (the depth of the valley 
at $\nu> 1$).  The phenomenon that clusters outnumber voids 
is reproduced by the simulations extremely well.  The 
amplitude of the genus curve 
agrees well between observations and simulations,
showing that the non-linear power spectrum and degree of phase correlation of the LRG
distribution smoothed at $R{_G} = 21 h^{-1}$ Mpc and 34$h^{-1}$Mpc are also modeled 
well.   
The small deviations of the genus curves from the Gaussian cases are 
quantified by our distortion parameters $\Delta\nu$, 
$A_V$, and  $A_C$.  These statistics for the observed LRG distribution 
are within approximately 1$\sigma$ of those expected from the 12 mock catalogs 
as shown in Figures 3 and 4. 

\begin{figure}
\epsscale{1.}
\plotone{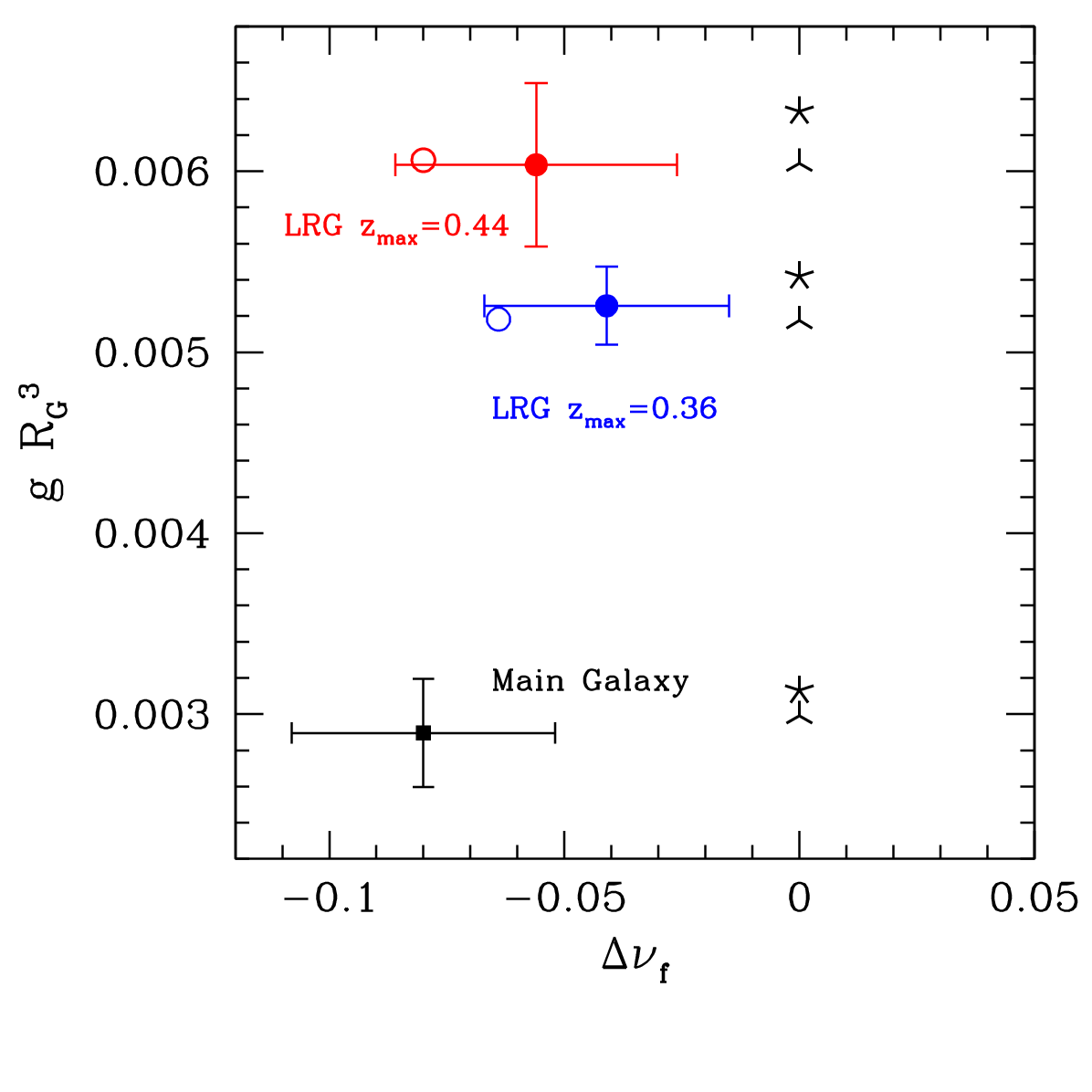}
\caption{
Genus per unit volume multiplied by the cube of the smoothing length: 
$g{R{_G}}^3$ versus the shift parameter $\Delta\nu$.
Filled circles with error bars are for the SDSS LRG galaxies in the SHALLOW sample (marked as
LRG $z_{\rm max}=0.36$) and in the DEEP sample. 
Open circles are the means of 12 mock SHALLOW and DEEP LRG surveys performed 
along the past light cone surface in a $\Lambda$CDM universe.
The filled square is from the SDSS galaxy distribution at 6 $h^{-1}$Mpc scale
taken from Park et al. (2005$a$).
The stars with 3 and 5 legs are the linear theory predictions for the WMAP 3 year
and 5 years cosmological parameters, respectively.
}
\end{figure}
\begin{figure}
\epsscale{1.}
\plotone{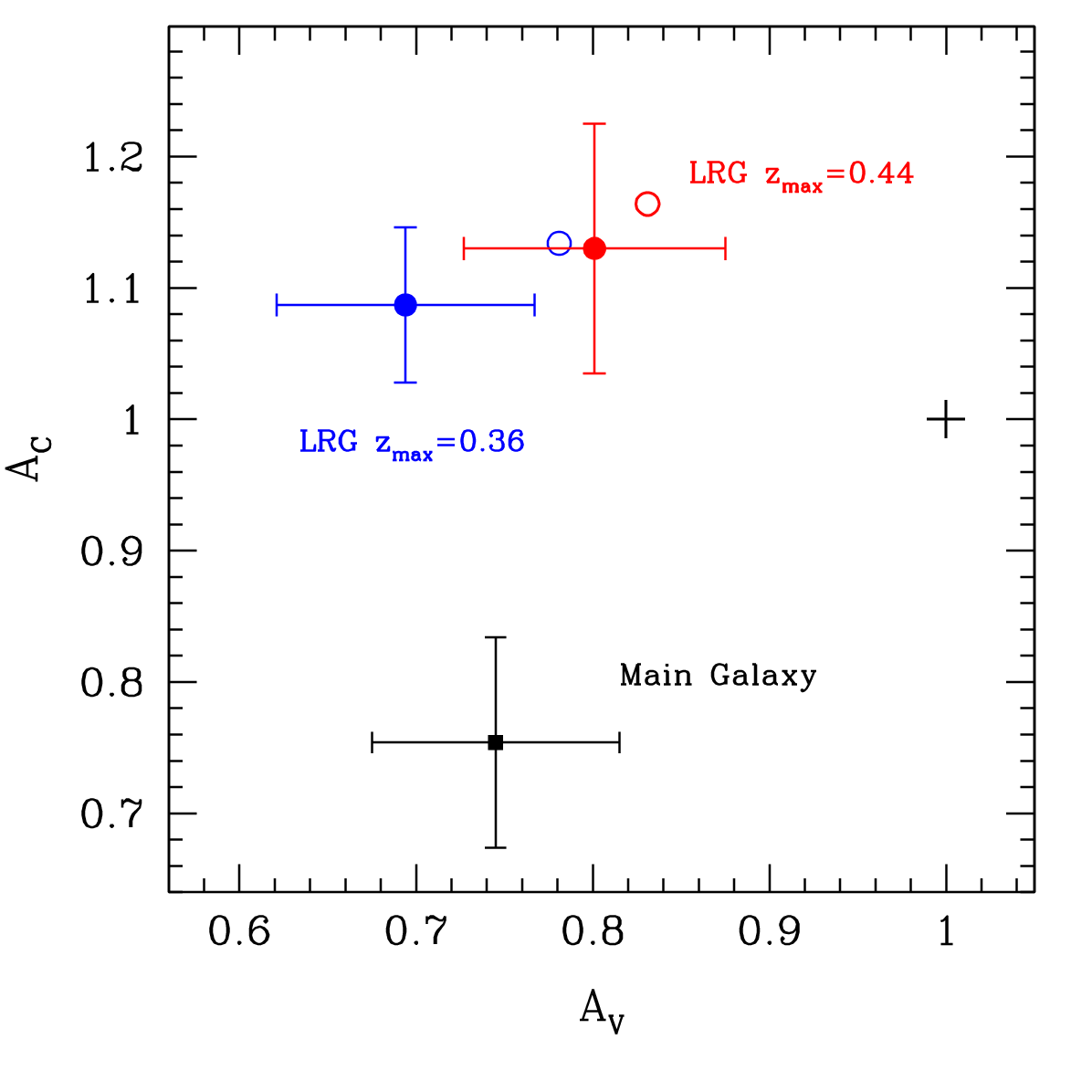}
\caption{Same as Fig. 3 but for the $A_C$ and $A_V$ parameters.
The cross on the right is the prediction for a random phase fluctuation.}
\end{figure}

\begin{deluxetable*}{lccccccccc}
\tablecolumns{10}
\tablecaption{Genus-Related Statistics}
\tablehead{
\colhead{Name}      &
\colhead{Magnitude} &
\colhead{Redshift}  &
\colhead{$\bar d$} &
\colhead{$R_{\rm G}$} &
\colhead{s}  &
\colhead{Amplitude}  &
\colhead{$\Delta\nu$} &
\colhead{$A_V$}&
\colhead{$A_C$}
}
\startdata
SHALLOW & $-21.2<M_{\rm g,0.3}<-23.2$& $0.2<z<0.36$& 22.1 & 21& 7 & $167.4 \pm 7.0$ &$-0.041 \pm 0.026$&$0.694\pm 0.065$ & $1.087  \pm 0.057$ \\
DEEP & $-21.8<M_{\rm g,0.3}<-23.2$& $0.2<z<0.44$& 35.0 & 34& 11 &$ 79.6  \pm 6.0$  &$-0.056  \pm 0.030$&$0.801 \pm 0.070$  & $1.127  \pm 0.092$\\
\enddata
\tablecomments{
Cols. (1) Name of samples,
(2) Rest-frame $g$-band absolute magnitudes,
(3) Redshift,
(4) Mean separation of galaxies in units of $h^{-1} \rm Mpc$,
(5) Smoothing length in units of $h^{-1}$Mpc,
(6) pixel scale in units of $h^{-1}$Mpc,
(7) The amplitude of the observed genus curve, 
(8) The shift parameter, 
(9) and (10) are the cluster and void abundance parameters, respectively.
Uncertainty limits are estimated from mock surveys in redshift space. 
}
\end{deluxetable*}

In Figure 3 the predictions for random phase curve (Eq. 2) 
are shown by stars with three and five legs for the cosmological
parameters estimated from
WMAP 3-year (Spergel et al. 2007) and 5-year maps only (Dunkley et al. 2008), 
respectively. 
In Figure 4 the random phase curve is indicated by a
point where $A_V = 1$ and $A_C = 1$.
The mock catalogs are indicated by open circles. 
The observational data are indicated by filled circles  with 1$\sigma$ 
error bars showing the $1\sigma$ variation from mock survey to mock survey.  
Importantly, there is no free fitting parameter here for any of these 
quantities.  So it is noteworthy that in $A, \Delta\nu, A_V,$ and $A_C$ 
the data are within about $1\sigma$ of the predicted values.  
The statistic showing the maximum difference between observation and simulation
is $A_V$; the observed voids are somewhat larger and fewer relative to those found
in mock SDSS LRG surveys. 

This suggests that finding the objects corresponding to LRGs 
is a relatively clean problem.
Our genus topology study shows that $N$-body simulation together with
identification of LRGs with the most massive dark subhalos can 
produce robust results in very good agreement with the observational data. 
Our results here show that the $\Lambda$CDM model we adopted in the
$N$-body simulation is a close approximation of the real universe.
The results also suggest that it will be possible to model well the formation 
of LRGs in the future Sloan III survey by using large $N$-body simulations to check 
and correct for any small systematic effects due to non-linear gravitational 
effects and biasing without needing free fitting parameters.  




\section{CONCLUSIONS}

We measured the 3D genus topology of large scale structure using the LRG 
galaxies in the Sloan Survey. The amplitude of the genus curve is now measured 
within about 4\% error at 21 $h^{-1}$Mpc smoothing scale. 
The results are consistent with the predictions of the $\Lambda$CDM
model with WMAP cosmological parameters, and with
Gaussian random 
phase initial conditions expected from standard inflation.  The topology is 
sponge-like.  Small distortions from the random phase curve expected from 
non-linear gravitational effects and biasing are observed.  
These are modeled 
well (at the $1\sigma$ level) by $N$-body simulations that pick the largest 
gravitationally bound halos not subject to tidal disruption as the sites for 
LRG galaxies to form.  The data on topology, including the number of voids 
and clusters, the location of the peak in the genus curve and its amplitude 
are all fit well without any free fitting parameters.  Thus,  the formation 
of LRGs appears to be a relatively clean problem that can be modeled well 
using cold dark matter and gravity alone, and this bodes well for using 
$N$-body simulations to calibrate measurement of  baryon oscillations using 
LRGs.    
 
\acknowledgments
JRG is supported by NSF grant AST-0406713.
CBP acknowledges the support of the Korea Science and Engineering
Foundation (KOSEF) through the Astrophysical Research Center for the
Structure and Evolution of the Cosmos (ARCSEC). YYC is 
supported by the WCU grant (No. R31-10016)
funded by the Korean Ministry of Education, Science and Technology.
The authors greatly thank Daniel J. Eisenstein for providing us with
the LSS-DR4plus Luminous Red Galaxy Sample.

Funding for the SDSS and SDSS-II has been provided by the Alfred P. Sloan Foundation, the Participating Institutions, the National Science Foundation, the U.S. Department of Energy, the National Aeronautics and Space Administration, the Japanese Monbukagakusho, the Max Planck Society, and the Higher Education Funding Council for England. The SDSS Web Site is http://www.sdss.org/.


{}
\end{document}